%% file: main.tex
\pdfoutput=1
\documentclass[conference]{IEEEtran}

\IEEEoverridecommandlockouts

\usepackage{amsmath}
\usepackage{amsfonts}
\usepackage{amssymb}
\usepackage{mathtools}
\usepackage{url}
\usepackage{cite}
\usepackage{epsfig}
\usepackage{xspace}
\usepackage{graphicx}
\usepackage{subfigure}
\usepackage{color}
\usepackage{stmaryrd}
\usepackage{multirow}
\usepackage{multicol}
\usepackage{relsize}
\usepackage{listings}
\usepackage{caption}
\usepackage{verbatim}
\usepackage{courier}
\usepackage{algorithm}
\usepackage{algpseudocode}

\newif\ifspace\spacefalse

\newtheorem{definition}{Definition}

\newcommand{\qed}{\hfill \ensuremath{\blacksquare}\\}

\graphicspath{{figs/}}

\newcommand\code[1]{\textsf{\small #1}}

\renewcommand{\baselinestretch}{0.95}
\begin{document}

\input{abs}
 \input{intro}
 \input{mot}
 \input{problem}

 \input{optimal}

 \input{design}
 \input{sim}
 \input{experiments}
 \input{related}
 \input{conclusion}

\bibliographystyle{plain}
\bibliography{bibliography}

\end{document}

%% file: abs.tex
\title{Power Redistribution for Optimizing Performance \\ in MPI Clusters}

\author{
\IEEEauthorblockN{Ramy Medhat}
\IEEEauthorblockA{Dept. of Elec. and Comp. Eng.\\
University of Waterloo, Canada \\
Email: \url{rmedhat@uwaterloo.ca}}
\and
\IEEEauthorblockN{Borzoo Bonakdarpour}
\IEEEauthorblockA{Department of Computing and Software \\ 
McMaster University, Canada\\
Email: borzoo@mcmaster.ca}
\and
\IEEEauthorblockN{Sebastian Fischmeister}
\IEEEauthorblockA{Dept. of Elec. and Comp. Eng.\\
University of Waterloo, Canada\\
Email: \url{sfischme@uwaterloo.ca}}
}

\maketitle

\begin{abstract}

Power efficiency has recently become a major concern in the high-performance 
computing domain. HPC centers are provisioned by a power bound which impacts 
execution time. Naturally, a tradeoff arises between power efficiency and 
computational efficiency. This paper tackles the problem of performance 
optimization for MPI applications, where a power bound is assumed. The paper 
exposes a subset of HPC applications that leverage cluster parallelism using 
MPI, where nodes encounter multiple synchronization points and exhibit 
inter-node dependency. We abstract this structure into a dependency graph, and 
leverage the asymmetry in execution time of parallel jobs on different nodes by 
redistributing power gained from idling a blocked node to nodes that are lagging 
in their jobs. We introduce a solution based on integer linear programming 
(ILP) for optimal power distribution algorithm that minimizes total execution 
time, while maintaining an upper power bound. We then present an online 
heuristic that dynamically redistributes power at run time. The heuristic shows 
significant reductions in total execution time of a set of parallel benchmarks 
with speedup up to $2.25$.

\end{abstract}


%% file: intro.tex
\section{Introduction}


Power efficiency in clusters for high-performance computing (HPC) and data 
centers is a major concern, specially due to the continuously rising 
computational demand, and the increasing cost and difficulty of provisioning 
enough power to satisfy such demand. This can be easily observed in the growing 
size of data centers that serve internet-scale applications. Currently, such 
data centers consume $1.3\%$ of the global energy supply, at a cost of $\$4.5$ 
billion. This percentage is expected to rise to $8\%$ by 
2020~\cite{koomey2011growth}. In fact, the rise in energy costs has become so 
prevalent that the cost of electricity for fours years in a data center is 
approaching the cost of setting up a new data center~\cite{barroso2005price}. 
This, consequently, results in HPC centers operating at tight power bounds, 
which in turn affects performance.

Extensive work has studied the trade-off between power efficiency and delay, 
sacrificing performance in favor of power~\cite{heath2005energy}. 
However, such sacrifice is intolerable in many applications, such as in HPC and 
cloud services that need to respect certain responsiveness features. On the 
contrary, there are approaches that target improving performance, which 
results in reduced energy consumption. This can be especially beneficial in 
heterogeneous clusters, which are rapidly becoming a favorable approach as 
opposed to homogeneous clusters~\cite{ramapantulu2014modeling}. Improved power 
efficiency, as well as lower cost and easier and cheaper upgrades are among the 
advantages of heterogeneous clusters. Reducing energy consumption is indirectly related to our objective of improving performance, yet there is very little work on optimizing power distribution of task dependency systems dynamically.

In this paper, we focus on improving the performance of MPI applications 
running on an HPC cluster, while satisfying a given power bound. To that end,
we present a novel solution to the problem that consists of the following:
\begin{itemize}
\item A technique to improve parallelism in MPI applications by {\em stretching} the execution of parts of the program that do not require immediate attention. This approach enables the execution of the stretched portion at a lower power level, providing more power to more critical tasks.
\item An ILP model that is capable of producing an optimal assignment of power to portions of the program.
\item An online power redistribution heuristic that dynamically transfers power from a blocked node to running nodes based on a priority ranking mechanism.
\item A novel implementation of an MPI wrapper that is capable of constructing a dependency graph online without the need to modify the code.
\end{itemize}


Our approach works as follows. We build an abstract model of a program 
executing in parallel as a set of program instances running on a cluster. To 
that end, we construct a dependency graph to encode the inter-dependency between 
blocks of non-synchronized code in the program instances, which we denote as 
jobs. The dependency graph allows us to identify the periods of time where a 
node will be blocked waiting on the output of other nodes. We exploit this 
behavior by {\em stretching} jobs so that we minimize the amount of time spent 
waiting for other nodes. Then, using the power constraints and execution time of 
jobs, we formulate our performance optimization problem as an integer 
linear program (ILP). The solution to this problem determines the 
optimum power-bounds to assign to jobs, such that the total execution time is 
minimized. We use this solution as a reference of goodness for the problem at 
hand.

\begin{figure}[h]
\centering
\includegraphics[width=1.0\columnwidth]{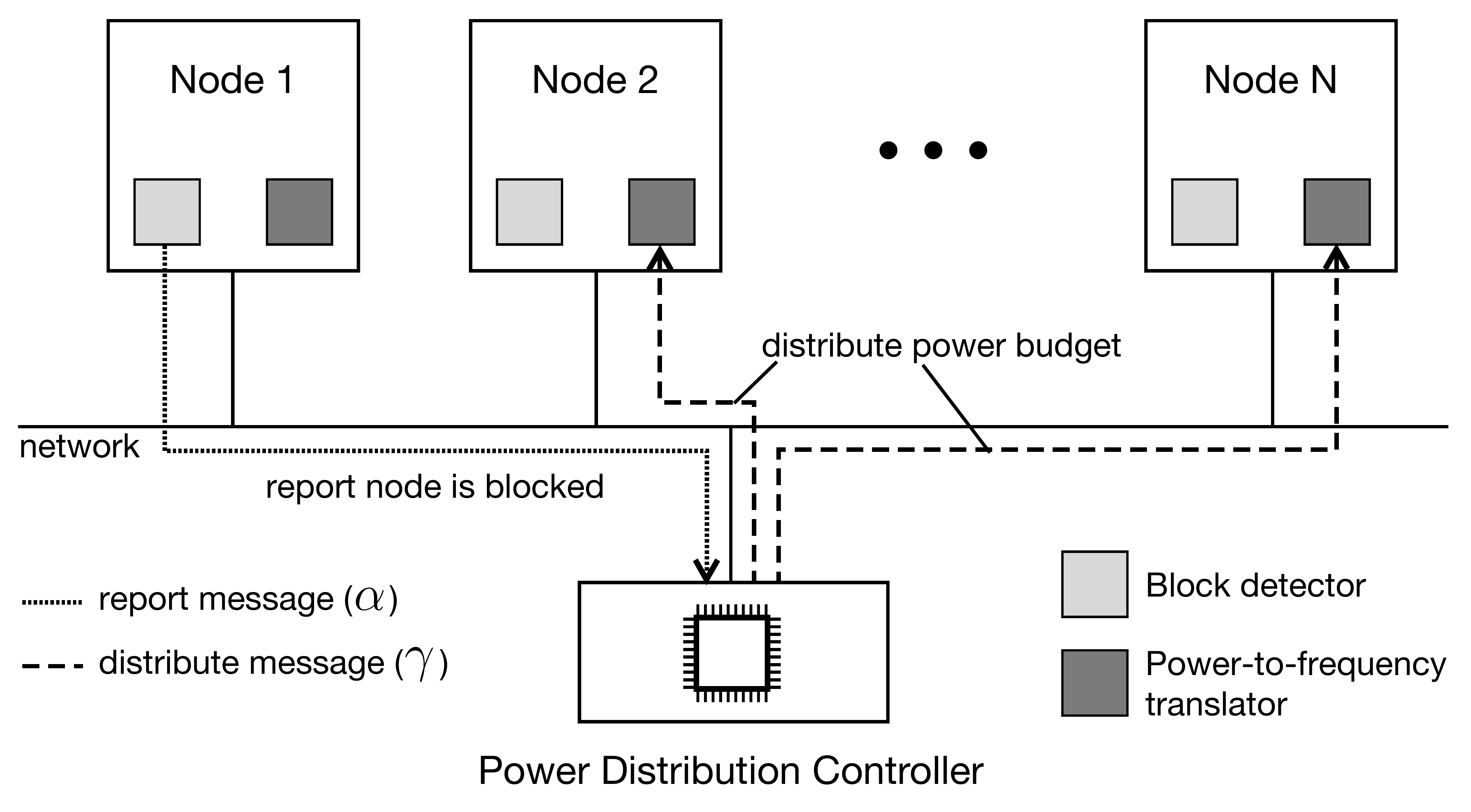}
\caption{Block diagram of an HPC cluster equipped with power distribution.}
\label{fig:topology}
\end{figure}

In order to deal with online power redistribution in MPI applications at run 
time, we introduce an online heuristic. Roughly speaking, the heuristic 
(see Figure~\ref{fig:topology}) detects when a node is blocked, and distributes 
its power (which will now drop to idle) over other nodes based on a priority 
ranking mechanism. We implement this heuristic as a wrapper for MPI, such that 
using it requires no modification to the program's source code. The wrapper is 
able to identify when a node is blocked, and the nodes responsible for blocking 
it. It communicates this information with a central power distribution 
controller, which in turn makes a decision on how to distribute the power gained 
by idling the blocked node.

A significant advantage of the proposed solution is that it is designed to improve performance of HPC applications with minimal requirements. Simply by linking existing code to the proposed MPI wrapper and connecting a low-end embedded board hosting the power distribution controller, a cluster can immediately benefit from dynamic power distribution.

To validate our approach, we conduct a set of simulations and actual 
experiments. First, we simulate a simple MPI program running using equal-share, 
ILP based, and heuristic based power distribution schemes. The simulation 
demonstrates the correlation between the variability in execution time and 
speedup. The improvement in execution time is promising, reaching a speedup of 
$2.5$ for ILP based distribution and $2.0$ for the heuristic based distribution. 
To demonstrate the applicability of the approach to homogeneous clusters, we run 
the simulation where all parallel jobs consume the same amount of time, and all 
nodes are identical. The speedup in this case is an encouraging $2.0$ for 
optimal distribution, and $1.64$ for the heuristic based distribution. 

We then experiment on a physical environment using two ARM-based boards that 
vary in CPU, operating system, and manufacturer. We implement our MPI wrapper 
that detects changes to node state and communicates it with a power distribution 
controller that executes our online heuristic. Using different benchmarks from 
the NAS benchmark suite, we demonstrate that our online heuristic can produce 
speedups up to $2.5$ times, only to be superseded by the optimal which reaches 
$2.78$ times. We also draw conclusions on the type of MPI program that would 
benefit most from our online power distribution heuristic.

\emph{Organization:} \ The rest of the paper is organized as follows.
Section~\ref{sec:motivation} introduces a motivating example to demonstrate our 
approach. Section~\ref{sec:prob} outlines a formal definition of the power 
distribution problem. Section~\ref{sec:optimal} presents an algorithm to achieve 
the optimal solution. Section~\ref{sec:design} details the design of our online 
heuristic. Section~\ref{sec:sim} presents the simulation results. 
Section~\ref{sec:exp} presents the MPI specific implementation and the 
experimental results. Finally, Section~\ref{sec:related} discusses related 
work, while we conclude in Section~\ref{sec:concl}.


%% file: mot.tex
\section{Motivation}
\label{sec:motivation}

This section outlines a motivating example to demonstrate the existence of an 
opportunity to optimize performance by redistributing power. 
Listing~\ref{lst:motivation} shows an abridged version of the \code{rank} 
function in the Integer Sort benchmark of the NAS Benchmark 
Suite~\cite{bailey1992parallel}. As can be seen in the code, there are $4$ main 
blocks of computation: The first block spans lines~\ref{rank_group1_start} 
to~\ref{rank_group1_end}. This is followed by a blocking collective operation at 
line~\ref{rank_mpi1}. This sequence continues until the last block spanning 
lines~\ref{rank_group4_start} to~\ref{rank_group4_end}. 

\begin{figure}[h]
\centering
\includegraphics[width=1\columnwidth]{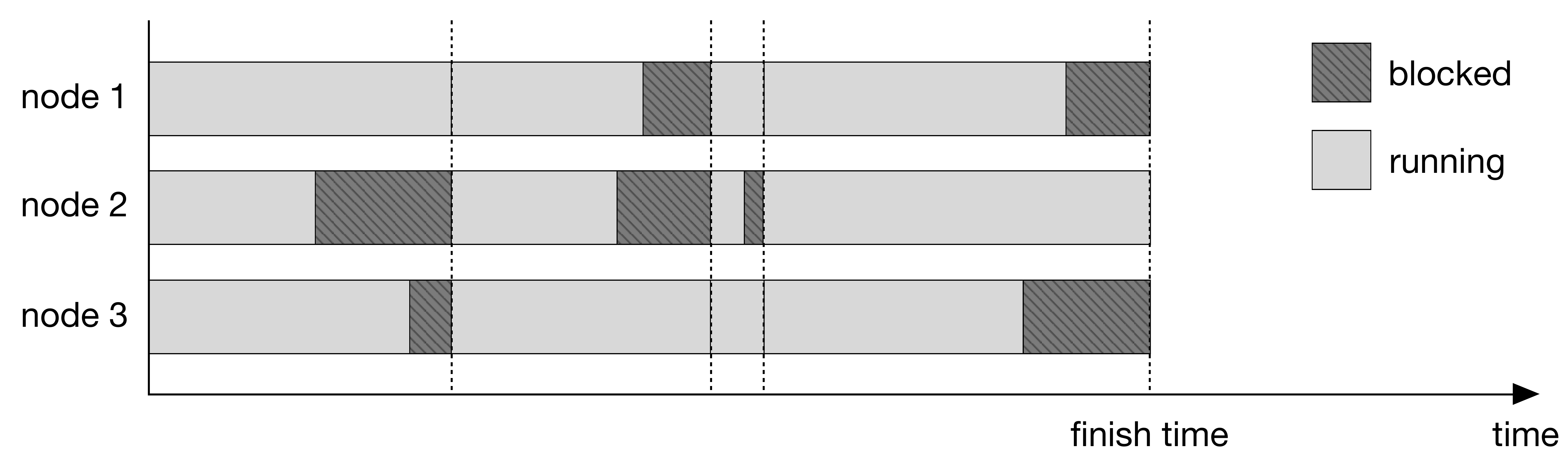}
\caption{A possible execution of \code{rank} on $3$ nodes.}
\label{fig:motivation}
\end{figure}

Figure~\ref{fig:motivation} illustrates a possible execution of this function 
on a $3$-node cluster. In this cluster, a maximum power bound is enforced, and 
this results in a power cap assigned per node, which limits its CPU frequency. 
However, it is possible and quite frequent that some nodes finish execution of a 
block of computation earlier than others, yet a blocking operation forces them 
to wait. This can be the result of, for instance, using heterogenous nodes, 
differences in workload, or nodes executing in different execution paths. This 
is demonstrated in Figure~\ref{fig:motivation} by the dark grey blocks in the 
figure, which we denote as {\em blackouts}. Naturally, execution cannot proceed 
until all nodes arrive at the {\em barrier}. This also applies to node-to-node 
send and receive operations.

\begin{figure}[h!]
\centering
\includegraphics[width=1\columnwidth]{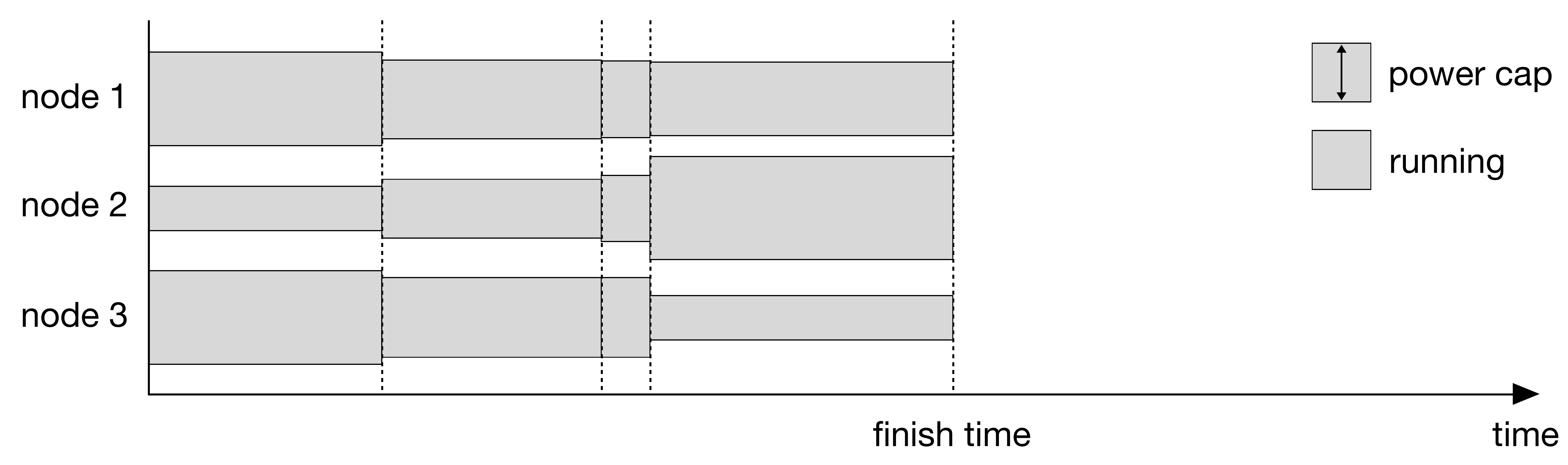}
\caption{An optimum execution of \code{rank} on $3$ nodes using power redistribution.}
\label{fig:motivation2}
\vspace{-5mm}
\end{figure}

Our research hypothesis is that an intelligent distribution of power can reduce 
these blackout periods, resulting in reduction of total execution time. This is 
demonstrated in Figure~\ref{fig:motivation2}. The thickness of a block 
indicates how much power it is allowed to consume. Blocks that consume a 
relatively short time in Figure~\ref{fig:motivation}, such as the first block in 
node 2, operate at a lower power cap in Figure~\ref{fig:motivation2} 
(demonstrated by reduced thickness). An optimum solution is capable of 
eradicating all blackouts, and minimizing those that are unavoidable (such as a 
ring send/receive). This results in a reduced total execution time within the 
cluster power bound.

\begin{lstlisting}[caption=Abridged \code{rank} method in the NPB IS benchmark, label={lst:motivation}][T!]
void rank( int iteration ) {
    for(i=0;i<NUM_BUCKETS+TEST_ARRAY_SIZE;i++) { (*@\label{rank_group1_start}@*)
    // Computation
    }
    for( i=0; i<TEST_ARRAY_SIZE; i++ ) {
    // Computation
    } (*@\label{rank_group1_end}@*)
    MPI_Allreduce( (*@$\cdots$@*) ); (*@\label{rank_mpi1}@*)
    for( i=0, j=0; i<NUM_BUCKETS; i++ ) { (*@\label{rank_group2_start}@*)
    // Computation
    } (*@\label{rank_group2_end}@*)
    MPI_Alltoall( (*@$\cdots$@*) ); (*@\label{rank_mpi2}@*)
    for( i=1; i<comm_size; i++ ) { (*@\label{rank_group3_start}@*)
    // Computation
    } (*@\label{rank_group3_end}@*)
    MPI_Alltoallv( (*@$\cdots$@*) );  (*@\label{rank_mpi3}@*)
    for( i=0; i<TEST_ARRAY_SIZE; i++ ) { (*@\label{rank_group4_start}@*)
    // Computation                     
    }  (*@\label{rank_group4_end}@*)
}   
\end{lstlisting}

%% file: problem.tex
\section{Formal Problem Description}
\label{sec:prob}

This section presents a formal description of the power distribution problem. 
Consider a set $\mathbb{N}=\{N_1,N_2, \dots, N_n\}$ of nodes in a parallel 
computing cluster that is required to satisfy a power bound $\mathbb{P}$. Each 
node runs a single instance of a parallel program. We model the execution of the 
program instance on a node as a sequence of jobs:
$$\mathcal{J}_i = \left< J_{i,1} J_{i,2} J_{i,j} \cdots \right> $$
where $J_{i,j}$ is the $j^{th}$ job on node $i$. A job represents a block of 
execution of the program instance on a single node that, once started, can be 
completed independently and without communication with other nodes. A job is 
defined as the following tuple:
$$J_{i,j} = \left< \tau(J_{i,j},P) , \theta(J_{i,j}), \pi(J_{i,j}) \right>$$
where
\begin{itemize}
\item $\tau(J_{i,j},P)$ is a function that encodes the execution time of job 
$j$ on node $i$ operating under power bound $P$, which directly enforces a 
maximum frequency that the CPU of node $i$ can utilize.

\item $\theta(J_{i,j})$ is a function that encodes the dependency of one job on 
a set of preoccuring jobs with the condition that it does not depend on multiple 
jobs in any other node. Such behavior can be expressed indirectly by chaining 
dependency. Naturally, in the serial execution of a program instance on node 
$i$, every job $j$ is dependent on its predecessor $j-1$. That is, $J_{i,j-1} 
\in \theta(J_{i,j})$. This implies that the execution of job $j$ cannot begin 
unless job $j-1$ is completed.

\item $\pi(J_{i,j})$ denotes the power bound that node $i$ should honour during 
execution of job $j$.
\end{itemize}

Our objective is to determine the mapping $\pi$ of all jobs on all nodes to 
their power bounds ($\pi(J_{i,j})$) such that:
\begin{enumerate}
\item The dependency of jobs $\theta$ is not violated;
\item The cluster power bound $\mathbb{P}$ is not exceeded, and
\item The total execution time is minimized.
\end{enumerate}

\subsection{Job Dependency Graph}

In order to calculate the total execution time, we construct a {\em job 
dependency graph}.\\

\begin{definition} [Job dependency graph]
\label{def:depgraph}
A {\em job dependency graph} $D$ is a directed acyclic graph, where vertices 
represent jobs $J_{i,j}$ and directed edges represent the dependency relation as 
described by $\theta$, such that if $(J_{i,j}, J_{i^\prime, j^\prime})$ is a 
directed edge of $D$, then $J_{i,j} \in \theta(J(i^\prime,j^\prime))$.\qed
\end{definition}

The job dependency graph is acyclic since a cycle will indicate circular 
dependency of jobs, which is impossible to occur since a job cannot be dependent 
on itself, directly or indirectly.

\subsection{Total execution time}

To discuss the total execution time, we first define the following:

\begin{itemize}

\item \textbf{Initial Job ($J^I$).} An initial job is a job that does not 
depend on any other job to begin execution, that is $\theta(J^I)=\{\}$. 
Normally an initial job is the first block of execution of a program instance on 
some node, up until the point of communication which is dependent on one or 
more other nodes. An initial job in a job dependency graph has no incoming 
edges.

\item \textbf{Final Job ($J^F$).} A final job is a job on which no other job 
depends. This is normally the final job to be executed by some node. A final job 
in a job dependency graph has no outgoing edges.
\end{itemize}

Given the notions of an initial job and a final job, we can now define an 
execution path.\\

\begin{definition} [Execution path]
\label{def:execpath}
An {\em execution path} $\rho$ is a path from an initial job to a final job 
($J^I \rightsquigarrow J^F$) in a job dependency graph. Clearly, such a path is 
a sequence of jobs: $$\rho = \left< J^I_{i,j}, J_{i^\prime,j^\prime}, \cdots , 
, J^F_{i^{\prime\prime},j^{\prime\prime}}\right>$$
such that every job in the sequence is dependent on its predecessor, that is $J_{i,j} \in \theta(J(i^\prime,j^\prime))$. We use the notation $\rho(l)$ to indicate the $l^{th}$ job in path $\rho$.
\qed
\end{definition}

The next step in preparing to calculate the total execution time is defining 
the scaling factor function $\mathcal{S}$. Let $\varrho = \{\rho_1, \rho_2, 
\rho_k, \cdots \}$ be the set of all execution paths in a job dependency graph 
$D$. The execution time $\epsilon$ of a path $\rho$ is calculated as follows: 
\begin{equation}
    \epsilon(\rho) = \sum_{l=0}^{|\rho|}{\tau(\rho(l), \pi(\rho(l))) }
\end{equation}
Thus, the execution time of a path 
is the sum of execution times of all the jobs in the path, given their 
respective power bounds. We can now define the total execution time.\\

\begin{definition} [Total execution time of a parallel program]
\label{def:totexectime}
Let a job dependency graph $D$ contain a set of execution paths $\varrho_D = 
\{\rho_1, \rho_2, \rho_k , \cdots\}$, and let the set $E_D = \{\epsilon(\rho_1), 
\epsilon(\rho_2), \epsilon(\rho_k), \cdots \}$ indicate the execution time of 
each path as mentioned above. We define the {\em total execution time} as $\mathbb{E}_D= \underset{k}{\text{max}} \left\{\epsilon(\rho_k)\right\}$ \qed
\end{definition}
Thus, the total execution time of a parallel program is the execution time of 
the longest execution path in the program's job dependency graph.

\subsection{Example of a Job Dependency Graph}

To demonstrate how a job dependency graph is constructed and how it is used to 
determine the total execution time of a parallel program, in this section, 
we introduce a simple MPI program as our running example throughout the paper. 
The program performs a set of commonplace MPI operations. The code in 
Listing~\ref{lst:example} demonstrates an MPI program that goes through 3 steps:
\begin{enumerate}
\item Broadcasts a message from the root node.
\item Sends a message between nodes in a ring.
\item Performs a reduction on a variable.
\end{enumerate}
Assume this program runs in a cluster of $3$ nodes. Based on the steps 
mentioned earlier, nodes will execute the following jobs:
\begin{itemize}
\item $J_{\_,1}$: represents lines~\ref{firstline}-\ref{bcast}. This is applicable to all nodes.
\item $J_{0,2}$ represents lines~\ref{secondline}-\ref{rootsend}. However, $J_{1,2}$ and $J_{2,2}$ represent lines~\ref{secondline}-\ref{otherrecv}.
\item $J_{0,3}$ represents line~\ref{rootrecv}, while the other two nodes represent line~\ref{othersend}.
\item $J_{\_,4}$ represents lines~\ref{thirdline}-\ref{reduce}.
\item $J_{\_,5}$ represents line~\ref{lastline}.
\end{itemize}

\begin{lstlisting}[caption=A simple MPI program., label={lst:example}, firstnumber=1][T!]
void main(int argc, char *argv[]) {
    int msg1, msg2, msg3; (*@\label{firstline}@*)
    int rank, size, next, prev;
    MPI_Init(&argc, &argv);
    MPI_Comm_rank(MPI_COMM_WORLD, &rank);
    MPI_Comm_size(MPI_COMM_WORLD, &size);
    
    if (rank == 0)
        msg1 = 10;
    
    MPI_BCast(&msg1,1,MPI_INT,0,MPI_COMM_WORLD);(*@\label{bcast}@*)
    
    next = (rank + 1) % size; (*@\label{secondline}@*)
    prev = (rank + size - 1) % size;
    
    if (rank == 0)
        msg2 = size;    
        
    if (rank == 0) {
        msg2--;
        MPI_Send(&msg2, 1, MPI_INT, next, rank, MPI_COMM_WORLD); (*@\label{rootsend}@*)
        MPI_Recv(&msg2, 1, MPI_INT, prev, prev, MPI_COMM_WORLD); (*@\label{rootrecv}@*)
    }
    else {
        MPI_Recv(&msg2, 1, MPI_INT, prev, prev, MPI_COMM_WORLD); (*@\label{otherrecv}@*)
        msg2--;
        MPI_Send(&msg2, 1, MPI_INT, next, rank, MPI_COMM_WORLD); (*@\label{othersend}@*)
    }
    
    msg2 = rank; (*@\label{thirdline}@*)    
    MPI_Reduce(&msg2, &msg3, 1, MPI_INT, MPI_MIN, 0, MPI_COMM_WORLD); (*@\label{reduce}@*)
    MPI_Finalize(); (*@\label{lastline}@*)
} 
\end{lstlisting}

\begin{figure}[h]
\centering
\includegraphics[width=0.7\columnwidth]{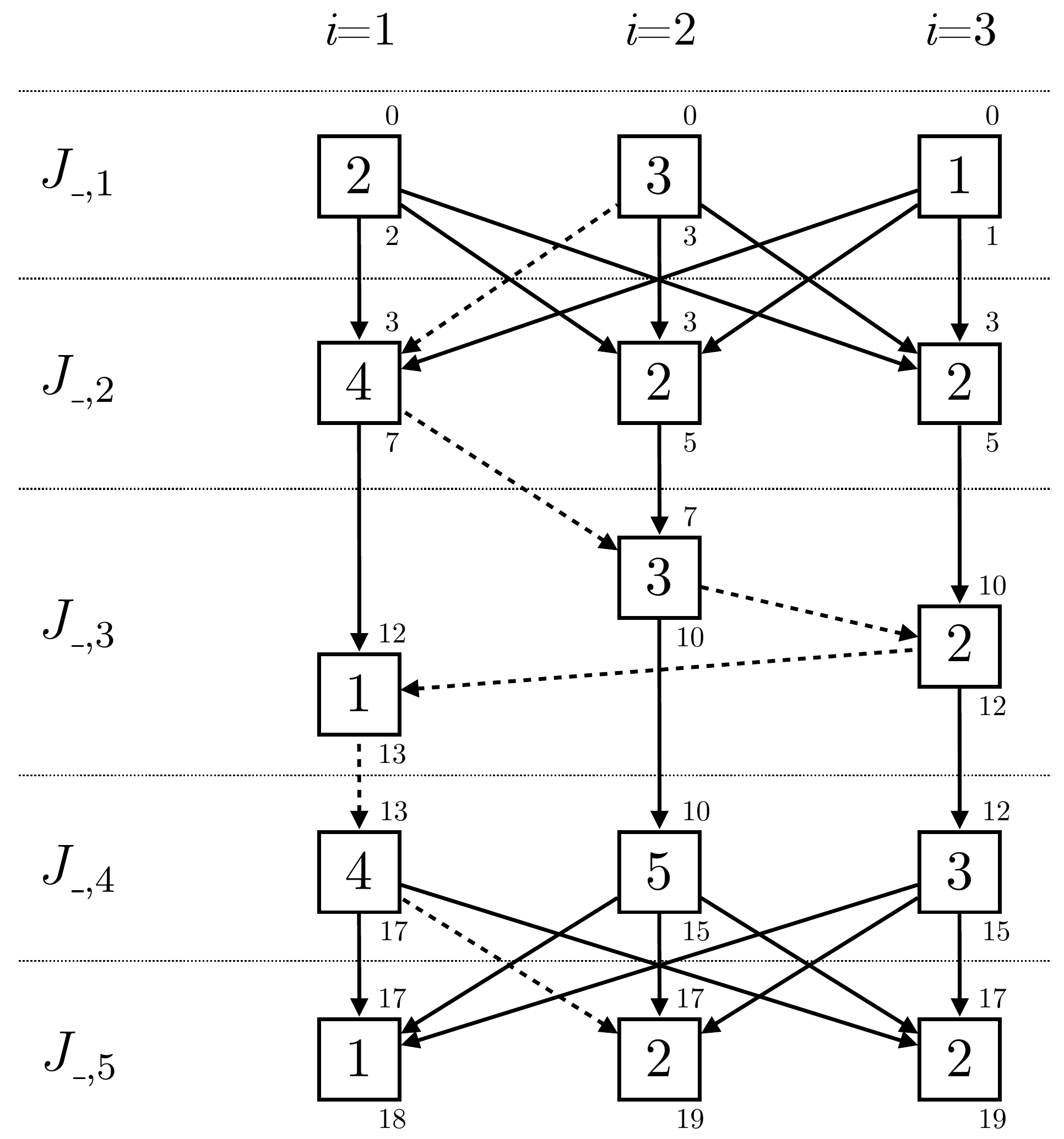}
\caption{Dependency graph of the program in Listing~\ref{lst:example}.}
\label{fig:depgraph1}
\end{figure}

In total, there are $15$ jobs in the system. Figure~\ref{fig:depgraph1} 
presents the dependency graph of the program with some hypothetical job 
execution times. These execution times are a result of applying the same power 
bound $\mathcal{P}$ on every node in the cluster, which we denote as the {\em 
nominal power bound}. The nominal power bound $\mathcal{P}$ is equal to 
$\mathbb{P} / N$, simply distributing the cluster power bound equally among all 
nodes in the system. Every block in the figure represents a job, which can be 
identified by its column, indicating to what node the job belongs to, and its 
row, indicating the index of that job in the sequence of node jobs. The nominal 
execution time of each job (that is $\tau(J_{i,j},\mathcal{P})$) is indicated by 
the number inside the block. The arrows represent dependency among nodes. As 
can be seen in the figure, the longest execution path starts with $J_{2,1}$ and 
proceeds along the dashed lines. Hence, the total execution time is $19$ time 
units.

Now, let us validate that the longest execution path is indeed indicative of 
the total execution time:
\begin{itemize}
\item  Execution starts at job $0$ in all nodes, which is a block of code that ends with a call to 
\code{MPI\_BCast}. A broadcast operation is an implicit barrier, and, hence, no 
node can proceed unless all $J_{\_,1}$ jobs are completed. This is visualized by 
connecting every $J_{\_,2}$ job with every $J_{\_,1}$ job.
\item Since $J_{2,1}$ takes the longest time, all $J_{\_,2}$ start after $3$ time units. This is indicated by the 
superscript of these blocks in the figure.
\item $J_{2,2}$, which ends with a call to \code{MPI\_Recv}, will block execution until node $1$ completes 
$J_{1,2}$ which ends with a call to \code{MPI\_Send}. Thus, $J_{2,3}$ is dependent on both its predecessor 
$J_{2,2}$ and $J_{1,2}$ which will send a message. The consequence of this dependency is that $J_{2,3}$ 
executes at $7$ time units, which is the maximum of the completion times of 
$J_{2,2}$ and $J_{1,2}$, as indicated 
by the subscript of the respective blocks. 
\item If we follow this process we can determine the completion time of all 
nodes, indicated by the subscript of the 
final jobs $J^F_{\_,5}$, after which the program terminates. The last jobs to complete are $J_{2,5}$ and $J_{3,5}$, which finish after $19$ time units.
\end{itemize}
As shown in Figure~\ref{fig:depgraph1}, the execution time of the longest path is the time at which all nodes in the cluster finish execution.

%% file: optimal.tex
\section{Optimal Solution}
\label{sec:optimal}


This section presents a method based on integer linear programming (ILP) 
to obtain the optimal solution for the power distribution problem. We, in 
particular, develop this method, so we have a reference of goodness for our 
online algorithm in Section~\ref{sec:design}.

In order to limit the number of variables in the ILP instance, we design an 
algorithm that establishes potential interleavings among jobs executing in 
different nodes. This algorithm is similar to real-time scheduling algorithms for task dependency on multiprocessors, with the added dimensions of job power bounds and variable execution times. The following subsection introduces the {\em Job Concurrency 
Optimization} algorithm.

\subsection{Job Concurrency Optimization Algorithm} 

The job concurrency optimization algorithm determines which jobs can execute 
concurrently without violating the dependency structure encoded in the job 
dependency graph. Since our objective is to reduce the length of blackouts, 
we can make an abstraction and avoid exploration of all possible interleavings 
in a parallel program. First, we begin by introducing the following 
definitions.\\

\begin{definition} [Job Max-Depth]
\label{def:maxdepth}
The max-depth $\delta$ of a job $J$ in a job dependency graph $D$ is the length 
of the longest path that starts from the initial job and ends with job $J$. 
That is 
$$\delta(J) = \max \left\{ l \mid \rho(l) = J  \; \wedge \; \rho 
\in \varrho_D\right\}
$$
\vspace{-4mm}
\qed
\end{definition}

\begin{definition} [Job Depth Range]
\label{def:depthrange}
The depth range $\Delta$ of a job $J$ in a job dependency graph $D$ is an integer interval defined as follows:
 $$\Delta(J) = [\delta(J),\beta(J)-1]$$ 
where $\delta(J)$ is the max-depth of job $J$ and the start of the interval, and 
$\beta(J)$ is the minimum max-depth of all $J$'s children, that is the set 
of jobs that are dependent on $J$:
$$\beta(J) ={\text{min}} \left\{ \delta(J^\prime) \mid J \in \theta(J^\prime)\right\}$$
where $J^\prime$ is a job in the job dependency graph.\qed
\end{definition}

\begin{figure}[h]
\centering
\includegraphics[width=1.0\columnwidth]{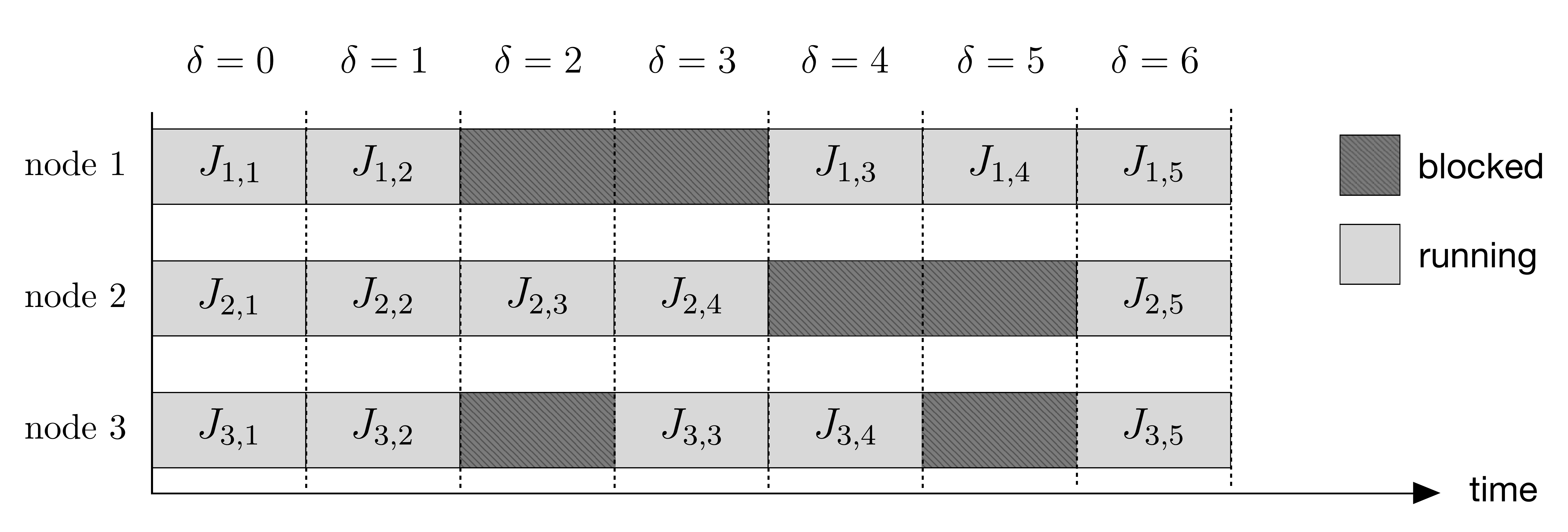}
\caption{Job concurrency as indicated by job max-depths.}
\label{fig:maxdepths}
\vspace{-2mm}
\end{figure}

Let us clarify the use of these definitions by referring to our earlier example 
in Listing~\ref{lst:example}, and the respective job dependency graph in 
Figure~\ref{fig:depgraph1}. Table~\ref{tab:maxdepths} shows the max-depths of 
all the jobs in the graph. Note that the max-depth is affected by the ring of 
sends/receives in job $3$ across all $3$ nodes. Figure~\ref{fig:maxdepths} 
visualizes how max-depths map to concurrency in execution. The dark grey blocks 
represent blackout periods, which should be optimally eradicated to reduce total 
execution time.

\begin{table}[h]
\caption{Max-depths of jobs in Figure~\ref{fig:depgraph1}.}
\label{tab:maxdepths}
\begin{center}
\begin{tabular}{l|ccc}
      & Node 1 & Node 2 & Node 3 \\ \hline
Job 1 & 0      & 0      & 0      \\
Job 2 & 1      & 1      & 1      \\
Job 3 & 4      & 2      & 3      \\
Job 4 & 5      & 3      & 4      \\
Job 5 & 6      & 6      & 6     
\end{tabular}
\end{center}
\vspace{-5mm}
\end{table}

\begin{figure}[h]
\centering
\includegraphics[width=1.0\columnwidth]{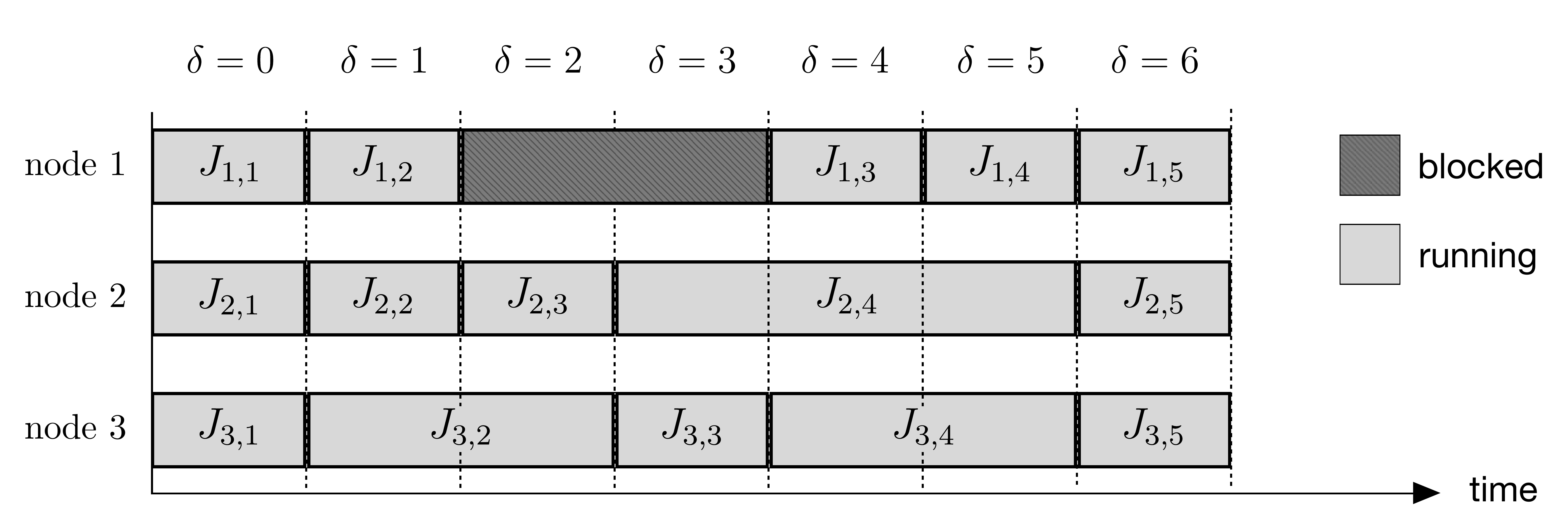}
\caption{Job concurrency after applying depth ranges.}
\label{fig:depthranges}
\vspace{-5mm}
\end{figure}

Table~\ref{tab:depthranges} shows the depth ranges of all jobs in the graph. The depth ranges 
allow us to revisit the job concurrency in Figure~\ref{fig:maxdepths} to produce an assignment 
that exhibits less blackouts. Figure~\ref{fig:depthranges} demonstrates applying depth ranges to 
determine optimum job concurrency. Observe job $J_{3,2}$ in the figure, which represents the code 
executed after the $\code{MPI\_BCast}$ and before the $\code{MPI\_Recv}$ called by the third node. 
The next block of code to be executed by node $3$ requires that node $2$ sends a message 
($J_{2,3}$). This implies that the execution of $J_{3,2}$ can be {\em stretched} beyond its max-depth level until node $2$ sends a message. {\em Stretching} a job in this manner implies allowing 
it to operate at a lower power level, thus enabling a higher power cap for other nodes in the 
cluster. 

\begin{table}[h]
\caption{Depth ranges of jobs in Figure~\ref{fig:depgraph1}.}
\label{tab:depthranges}
\begin{center}
\begin{tabular}{l|ccc}
      & Node 1 & Node 2 & Node 3 \\ \hline
Job 1 & [0,0]      & [0,0]      & [0,0]      \\
Job 2 & [1,1]      & [1,1]      & [1,2]      \\
Job 3 & [4,4]      & [2,2]      & [3,3]      \\
Job 4 & [5,5]      & [3,5]      & [4,5]      \\
Job 5 & [6,6]      & [6,6]      & [6,6]     
\end{tabular}
\end{center}
\vspace{-8mm}
\end{table}

Thus, Figure~\ref{fig:depthranges} shows that utilizing depth ranges helps 
remove blackout periods in execution, except for unavoidable blackouts such as a 
message ring as shown in our example.

The algorithm is simple to implement, since calculating the max-depths of jobs 
in the dependency graph is a straight-forward node traversal, and the complexity 
is $O(E)$, where $E$ is the number of edges in the graph. Finding max-depth in 
the case of the job dependency graph is thus linear in the size of the graph. 
Likewise, computing depth ranges requires iterating over the outgoing edges of 
every job in the graph, resulting in similar complexity.

\subsection{ILP Instance}

\begin{figure}[h]
\centering
\includegraphics[width=1.0\columnwidth]{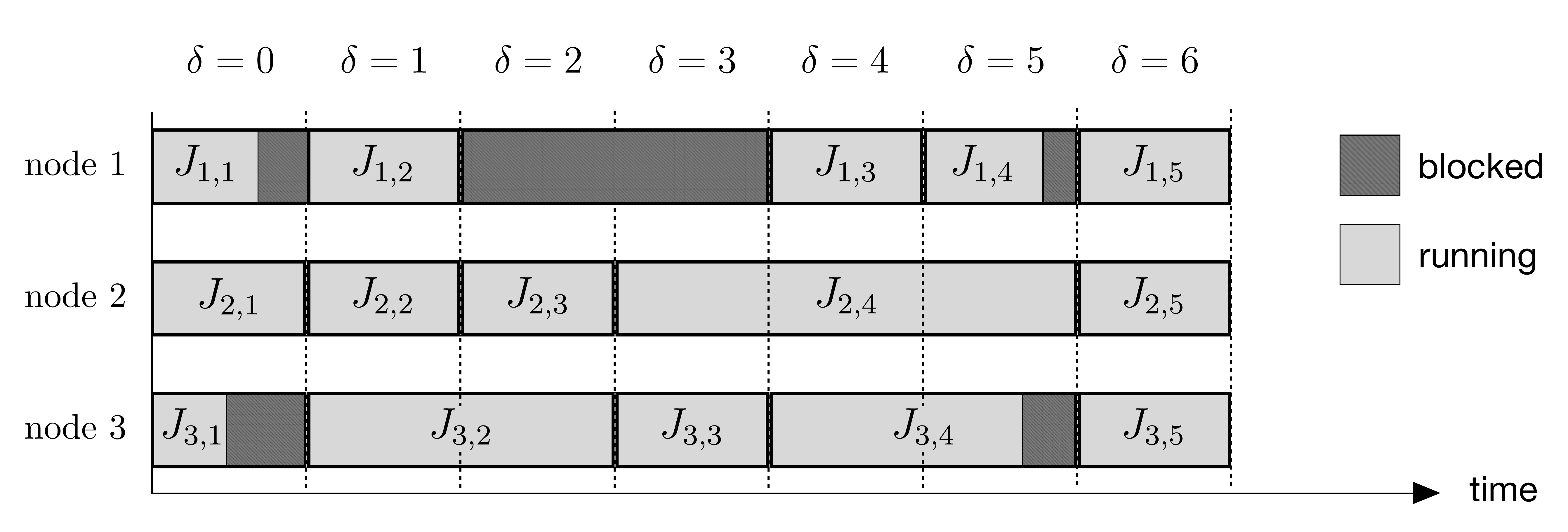}
\caption{Blackouts in execution due to difference in execution time.}
\label{fig:ilp}
\vspace{-3mm}
\end{figure}

This section introduces the ILP instance used to find the optimum power bound 
assignment $\pi$, which assigns a power bound to every job in the dependency 
graph. Refer to Figure~\ref{fig:depthranges}, which shows how depth ranges 
reduce blackouts. Note that the figure does not represent the actual execution 
times of all jobs. For instance, the execution time of jobs $J_{\_,1}$ in 
Figure~\ref{fig:depgraph1} are $2$, $3$, and $1$, respectively. This implies a 
blackout will exist at node $1$ between $2$ and $3$ time units, and at node $3$ 
between $1$ and $3$ time units. Figure~\ref{fig:ilp} illustrates how such 
blackouts would occur.  

Solving the ILP instance produces a set of power-bound assignments to jobs that 
minimizes these blackout periods. The ILP instance is detailed as follows.\\

\noindent \textbf{Variables}. The ILP instance abstracts the range of power 
bounds that can be applied to a certain node into a finite set of power bounds 
that map to operating frequencies of the node's CPU. This is a reasonable 
abstraction since any CPU supports a finite set of operating frequencies, and we 
can hypothetically determine the power ceiling of the node when operating at 
every respective frequency. Hence, we introduce the following variables:
\begin{itemize}

\item {\em (Job-to-power-bound-assignment $x_{j,b}$)} This is a binary 
variable that indicates whether job $j$ is assigned to power bound $b$. In a 
dependency graph consisting of $10$ jobs, where each job can operate at $5$ 
different power bounds, the ILP model will contain $50$ 
job-to-power-bound-assignment variables.

\item {\em (Maximum execution time $t$)} This variable represents the maximum 
time it takes any node in the system to finish execution.
\end{itemize}

\noindent \textbf{Constraints}. The model employs $3$ types of constraints.
\begin{itemize}

\item {\em (Unique power bound assignment)} These constraints ensure that no 
job is assigned to two different power bounds. There is one such constraint per 
job in the graph.
$$\forall j : \sum_{b}{x_{j,b}} = 1$$

\item {\em (Cluster power bound enforcement)} These constraints enforce the 
power bound $\mathbb{P}$ on the entire cluster. Generating these constraints 
relies on the output of the job concurrency optimization algorithm (refer to 
Figure~\ref{fig:depthranges}). Every depth level column indicates which jobs 
will execute concurrently. For instance, job $J_{3,2}$ is concurrent with 
$\{J_{2,2},J_{1,2}\}$ at depth level $\delta = 1$. It is also concurrent with 
$J_{2,3}$ at depth level $\delta = 2$. Hence, there is one power bound 
enforcement constraint per depth level in the graph:
$$\forall \delta : \sum_{\delta_j}{x_{j,b} \times b} \le \mathbb{P}$$
where $\delta_j$ is a set that contains any job for which $\delta$ is within 
its depth range:
$$\delta_j = \{J \mid \delta \in \Delta(J)\}$$

\item {\em (Maximum execution time)} These constraints ensure that no node 
executes beyond the maximum execution time variable $t$, which is the variable 
to be minimized.
$$\forall i: \sum_{j \in \mathcal{J}_i}{\sum_{b}{x_{j,b} \times \tau(j,b)}} \le t$$
where $i$ indicates the node, $\mathcal{J}_i$ is the sequence of jobs in that node, $j$ is a job in that sequence, and $\tau(j,b)$ is the execution time of job $j$ under power bound $b$.
\end{itemize}
The total number of constraints in the model is the result of the following formula:
$$\sum_{i}{|\mathcal{J}_i|} + \underset{J}{\text{max}} \{\delta(J)\} + n$$
where $n$ is the number of nodes.\\

\noindent \textbf{Optimization objective}. Finally, the objective of the model 
is to minimize the maximum execution time:
\begin{equation}
\min\; t
\end{equation}

%% file: design.tex

\section{Online Heuristic for Power Redistribution}
\label{sec:design}

This section introduces the design of an online heuristic that dynamically 
distributes power. We approach the design of the heuristic with the following 
set of objectives:
\begin{itemize}
\item The optimum solution introduced in the previous section requires the 
knowledge of the execution time of every job at every CPU frequency. This is not 
realistically available for a running application, and, hence, obtaining an 
optimum online solution is not possible. Thus, our objective is to build an 
algorithm that receives realistic input and can make online decisions.

\item Making power distribution decisions must incur minimal overhead; i.e., in 
a thrashing-free manner.

\item Design the algorithm to be lightweight, executable on non-sophisticated 
power-efficient hardware.
\end{itemize}

The heuristic targets HPC clusters composed of heterogenous nodes. 
Figure~\ref{fig:topology} illustrates the structure of such a cluster, and 
introduces the following components:

\begin{itemize}
\item \textbf{Block detector.} The block detector is responsible for detecting 
when a node becomes blocked, awaiting some input from one or more other nodes. 
It is also responsible for detecting when the node becomes active again. it 
reports these changes in state to the power distribution controller. The block 
detector is further explained in Subsection~\ref{subsec:report}.

\item \textbf{Power distribution controller.} The power distribution controller 
receives messages whenever a node is blocked or unblocked. Since the blocked 
node will transition to idle, the total power consumption of the cluster will 
drop. This will provide a {\em power budget} that can be distributed to other 
nodes. The power distribution controller makes a decision on how to distribute 
the power budget on {\em running} nodes. The decision procedure is the core of 
our online heuristic, which is explained in detail in 
Subsection~\ref{subsec:algorithm}.

\item \textbf{Power-to-frequency translator.} This component receives the 
distribute message and translates the power bound dictated by the power 
distribution controller to a CPU frequency. It selects the maximum CPU 
frequency that can satisfy the power bound in the message and forces the node 
to operate at that frequency.
\end{itemize}

\subsection{Block Detector}
\label{subsec:report}

As shown in Figure~\ref{fig:topology}, the block detector sends a report 
message to the power distribution controller whenever a change in the node 
state is detected. A report message is a tuple defined as follows: 
$$\alpha=\left<s,i,B,p_g\right>$$
where

\begin{itemize}
\item $s$ is the state of the node, whether \code{Blocked} or \code{Running}.

\item $i$ is the index of the node from which the report message originated.

\item $B$ is a set of node indices that are causing node $i$ to be blocked. If 
$s = \code{Running}$, $B$ becomes an empty set.

\item $p_g$ is the power gained by blocking node $i$, which is calculated as 
follows:
$$p_g = p_{f_c} - p_s$$

where $f_c$ is the CPU frequency before the block is encountered, $p_{f_c}$ is 
the power consumed by running the CPU at frequency $f_c$, and $p_s$ is the 
idle power.
\end{itemize}

To compute $p_g$, we require that each node hosts a lookup table mapping CPU 
frequency to power. This is obtained by executing a simple benchmark that loads 
the CPU $100\%$ at each frequency, and records the power consumption. However, 
if a node hosts a multicore CPU and is executing multiple jobs concurrently, one 
per core, the power gain becomes the current power minus the power consumed when 
one less core is active. Hence, we require that the lookup table includes the 
power consumption of the node at each available frequency {\em and} at every 
possible number of active cores. For instance, a quad-core CPU that supports 
$10$ different frequencies will result in a lookup table of $40$ entries. An 
entry will be identified by (1) the number of active cores in parallel (e.g., 
$1-4$ for quad-core CPUs), and (2) the CPU frequency.

To formally define this, let $p_{m,f}$ be the power consumption of the node 
when $m$ cores are active and the CPU frequency is $f$. Let $m_c$ and $f_c$ be 
the number of active cores and  the CPU frequency before the block is 
encountered on a job executing in one core. In that case, $p_g$ is calculated as 
follows:
\begin{equation}
\label{eq:pg}
p_g=p_{(m_c-1,f_c)} - p_s
\end{equation}

\subsection{Power Distribution Controller -- Online Heuristic Design}
\label{subsec:algorithm}

Algorithm~\ref{alg:heuristic} details the logic behind the power distribution 
heuristic. The heuristic is initialized with a cluster power bound $\mathbb{P}$ 
watts and an empty (online) dependency graph $G = (V, E)$. The following steps 
detail the operation of the heuristic:

\begin{enumerate}
\item The function {\sc ProcessMessage} is invoked whenever the power 
distribution manager receives a report message $\alpha$ from any node in the 
cluster (line~\ref{line:processmessage}).

\item Lines~\ref{line:upd_graph_start}-\ref{line:upd_graph_end} in function {\sc 
ProcessMessage} update the online dependency graph using the received message. 
{\sc ProcessMessage} creates a vertex for the node if it does not already exit, 
and connects the vertex to other vertices representing nodes that are blocking 
the sender node.

\item Lines~\ref{line:power_budget_start}-\ref{line:power_budget_end} calculate 
the available power budget by adding the power gain $p_g$ of all blocked nodes 
in the graph.

\item The function then calls {\sc RankGraph} which calculates the priority of a 
node based on the number of other nodes that it is blocking. A node of rank $0$ 
has no incoming edges, and hence is not blocking any node. A node of rank $1$ 
blocks one other node, and so on.

\item Finally, function {\sc DistributePower} is responsible for distributing 
the power budget over running nodes. If a node is blocking two other nodes, it 
will receive twice the amount of power received by a node that is blocking only 
one other node. This strategy allows the system to gradually increase the power 
bound of the older blocking nodes, since every time any node is blocked, the 
older blocking node receives a portion of the power gain.
\end{enumerate}

\input{Algorithm1.tex}

%% file: Algorithm1.tex
\algrenewcommand\alglinenumber[1]{\small #1:}
\begin{algorithm}
\footnotesize
\begin{algorithmic}[1]
\State INPUT: Cluster power bound $\mathbb{P}$, number of nodes $n$
\State declare $G=(V,E)$, initially $V = E = \{\}$ \Comment{Online Dependency 
Graph}
\State declare $p_o = \mathbb{P} / n$
\Function{ProcessMessage}{$\alpha$} \label{line:processmessage}   
    \If{$\alpha.i \not\in V$} \label{line:upd_graph_start}
        \State $v \gets$ \Call{AddVertex}{$V$, $\alpha.i$}
    \Else
        \State $v \gets V[\alpha.i]$ 
    \EndIf
    \State $v.s \gets \alpha.s$
    \State $v.p_g \gets \alpha.p_g$
    \State \Call{UpdateEdges}{$G$, $v$, $\alpha.B$} \Comment{Update connections of $v$ using $\alpha$} \label{line:upd_graph_end}
    
    \State declare $\varepsilon = 0$ \Comment{Power budget}  \label{line:power_budget_start}
    \For{$u \in V$}
    \If{$u.s = \text{Blocked}$}
        \State $\varepsilon \gets \varepsilon + u.p_g$
    \EndIf
    \EndFor   \label{line:power_budget_end}
    
    \State $t \gets$ \Call{RankGraph}{}
    
    \State \Call{DistributePower}{$\varepsilon$, $t$}
\EndFunction

\vspace{2mm}

\Function{UpdateEdges}{$G$, $v$, $B$}
    \State \Call{ClearOutgoingEdges}{$v$}
    \For{$u \in B$}
        \State \Call{AddEdge}{$E$, $v$, $u$}
    \EndFor
\EndFunction

\vspace{2mm}

\Function{RankGraph}{} \label{line:rankgraph}
    \State $t \gets 0$
    \For{$u \in V$}
    \If{$u.s = \text{Running}$}
        \State $u.r \gets |\{e=(a,b) \in E \mid e.b = u\}|$
        \State $t \gets t + u.r$ \Comment{Sum of all ranks}
    \EndIf
    \EndFor
    \State {\bf return} $t$
\EndFunction

\vspace{2mm}

\Function{DistributePower}{$\varepsilon$, $t$} \label{line:distributepower}
    \For{$u \in V$}
    \If{$u.s = \text{Running}$}
        \State $p_b^\prime = p_o + \varepsilon \times u.r/t$
        \If{$u.p_b \neq p_b^\prime$}
        \State $u.p_b = p_b^\prime$
        \State $\gamma \gets (u.i,u.p_b)$
        \State \Call{SendPowerBound}{$\gamma$}
        \EndIf
    \EndIf
    \EndFor
\EndFunction

\end{algorithmic}
\normalsize
\caption{Power distribution online heuristic.}
\label{alg:heuristic}
\end{algorithm}

%% file: sim.tex
\section{Simulation results}
\label{sec:sim}
\vspace{-2mm}
To validate the intuition behind our model, the ILP solution, and the online 
heuristic, we implement a simulator to calculate the total execution time of an 
MPI program. The simulator is initialized with the following:
\begin{itemize}
\item a text file detailing the job dependency graph,
\item a cluster power bound, and
\item the type of simulation: {\em Equal-share}, {\em ILP}, or {\em Heuristic}.
\end{itemize}

The {\em Equal-share} simulation assigns equal power bounds to all nodes in the 
cluster. The {\em ILP} simulation first solves the power assignment problem for 
an optimal (or nearly optimal due abstractions) solution, and then simulates 
execution using the resulting job-to-power-assignments. The {\em Heuristic} 
simulation applies the online power distribution algorithm.

\begin{figure}[t]
\centering
\includegraphics[width=0.8\columnwidth]{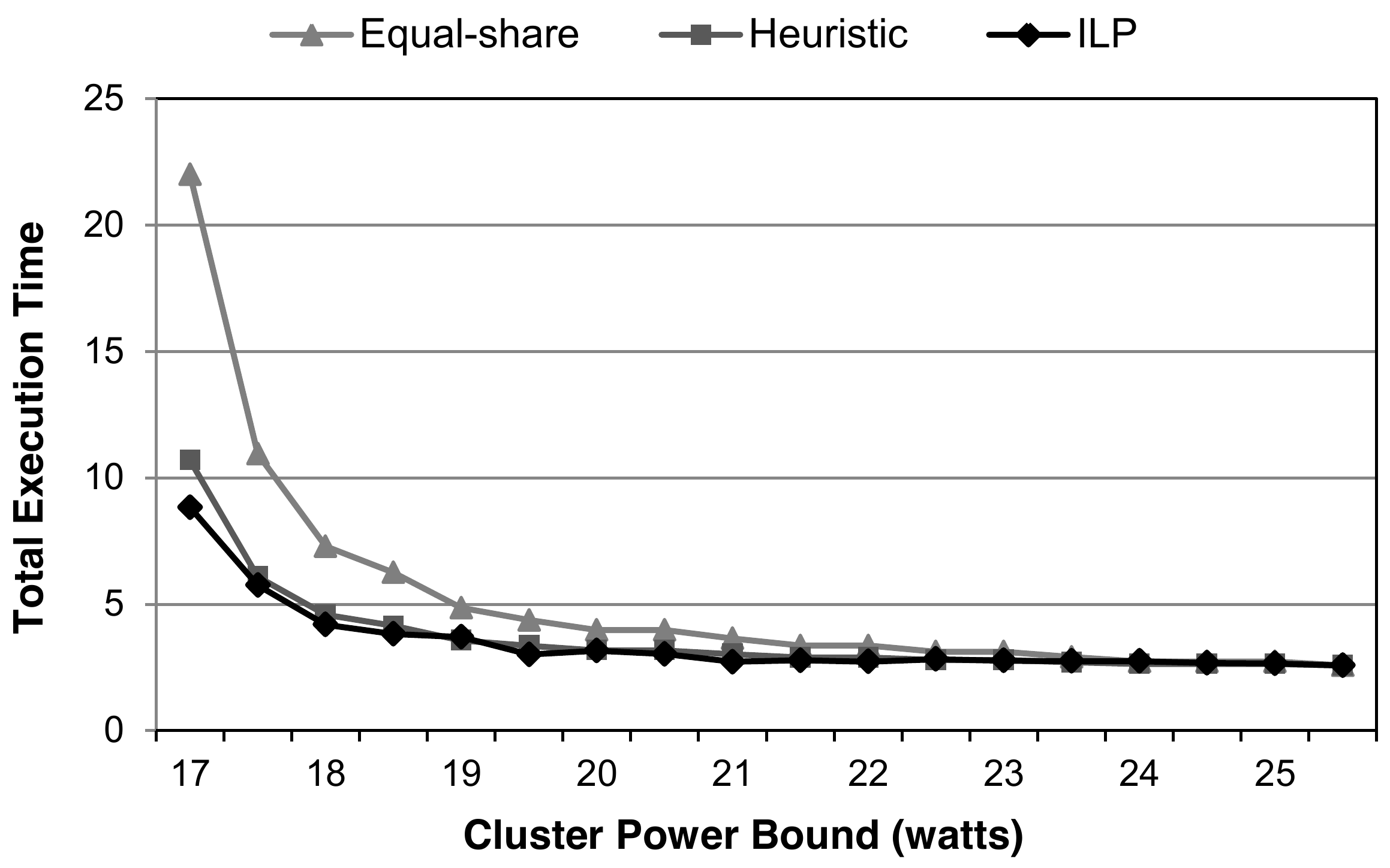}
\caption{Simulation results of the dependency graph in Figure~\ref{fig:depgraph1}.}
\label{fig:sim}
\vspace{-3mm}
\end{figure}

Figure~\ref{fig:sim} shows the results of simulating the dependency graph in 
Figure~\ref{fig:depgraph1}. The power-to-frequency lookup values, as well as the 
execution time of jobs at different CPU frequencies are measured on an Arndale 
Exynos 5410 ARM board. The results indicate that the ILP-based solution excels 
at the lower power bounds, producing a $2.5$ speedup versus equal-share. The 
heuristic also produces a significant speedup of $2$ at lower power bounds. The 
improvement for both ILP-based solution and the heuristic decreases until it 
matches the execution time of equal-share as the power bound is relaxed.  This 
is expected since at a relaxed power bound the nodes are already operating at 
their maximum frequencies.

These results are based on the assigned execution times in the job dependency 
graph in Figure~\ref{fig:depgraph1}, which are completely synthetic. To add some 
notion of ground truth to the simulations, we rerun the simulation given that 
the execution times of all jobs is the same. Hence, no bias exists that would 
favor a power distribution alternative to equal-share. In such a case, the 
ILP-based solution still outperforms equal-share at lower power bounds, 
producing a speedup of $2$, while the heuristic speedup is $1.64$. The 
improvement comes from the fact that the ring communication pattern forces 
blocking in the equal-share distribution, even when the execution times of jobs 
is the same. This is improved significantly by applying the {\em stretching} of 
jobs across multiple depth levels, and distributing power optimally on running 
nodes.

\begin{figure}[t]
\centering
\includegraphics[width=0.8\columnwidth]{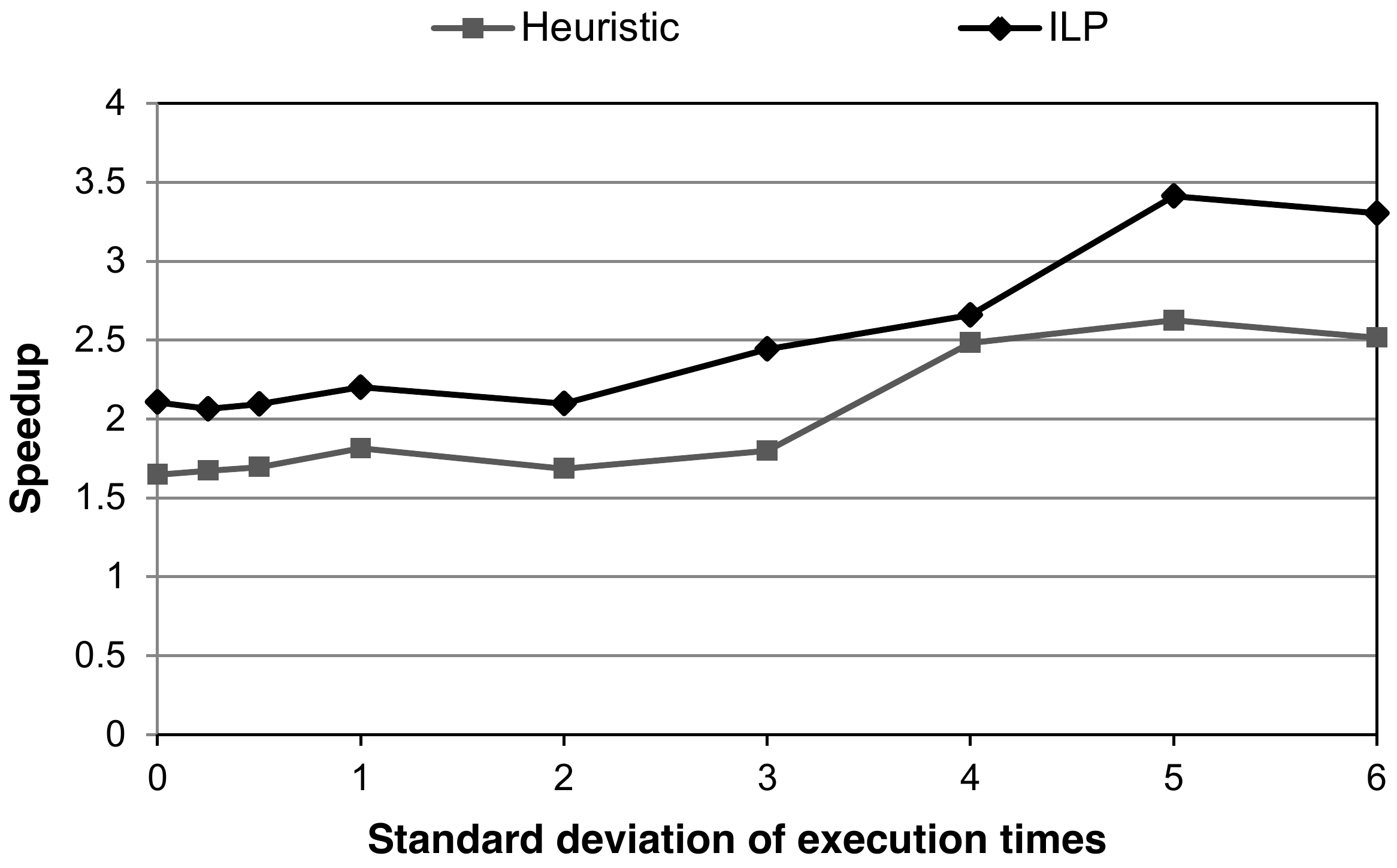}
\caption{Simulation results using different standard deviations of execution times.}
\label{fig:sim2}
\vspace{-1mm}
\end{figure}

In light of these results, we construct a set of experiments based on the same 
dependency graph, yet varying in execution times. We quantify the variation in 
execution times using the standard deviation of execution times of individual 
jobs. Hence, the experiments present synthesized execution times to target 
specific standard deviations. The standard deviation starts at $0$ and increases 
till $6$, given a mean of $10$ time units. Figure~\ref{fig:sim2} illustrates the 
speedup gained by running the heuristic and the ILP solution, given the minimum 
possible cluster power bound. The figure shows a trend of increasing speedup as 
the variation increases, which confirms our intuition that our algorithms excel 
when execution times exhibit more variability. Yet, at high variability, speedup 
becomes unstable since it is heavily dependent on the specific execution times 
assigned to jobs. 

%% file: experiments.tex
\section{Implementation and Experimental Results}
\label{sec:exp}


\subsection{Implementation}
\label{sec:implementation}

Although the model proposed in the paper can generally map to clusters that 
exhibit task dependency, we focus our implementation on MPI clusters. To that 
end, we implement an MPI wrapper with underlying logic to perform the 
functionality of the Block Detector (see section~\ref{subsec:report}). The power 
distribution controller is implemented as a standalone lightweight UDP server 
that receives report messages and responds with distribute messages.

\subsubsection{MPI wrapper}
The MPI wrapper is designed to intercept MPI calls and deduce whether the node will be blocked or unblocked. Currently the wrapper supports \code{MPI\_Send}, \code{MPI\_Recv}, \code{MPI\_BCast}, \code{MPI\_Wait}, \code{MPI\_Scatter}, \code{MPI\_Reduce}, and \code{MPI\_AlltoAll}.

The wrapper uses the parameters of the MPI call to deduce the nodes that are 
blocking the current node, and then creates a report message and transmits it to 
the power distribution controller. Listing~\ref{lst:mpibcast} shows an abridged 
version of the \code{MPI\_BCast} wrapper. The function \code{power\_gain()} in 
line~\ref{line:powergain} calculates the power gain according to 
Equation~\ref{eq:pg}. Function \code{all\_other\_nodes()} in 
line~\ref{line:other_nodes} returns the identifiers of all nodes in the cluster 
with the exclusion of the current node. Since the operation is a broadcast, the 
current node will in fact not proceed with execution until all other nodes in 
the cluster have reached the same \code{MPI\_BCast} call.

\begin{lstlisting}[caption=\code{MPI\_BCast} wrapper., label={lst:mpibcast}, firstnumber=1][T!]
int pMPI_Bcast(void *buffer, int count, MPI_Datatype type, int root, MPI_Comm comm)
{
    report_message msg;
    msg.node_id = my_rank;
    msg.state = Blocked;
    msg.power_gain = power_gain(); (*@\label{line:powergain}@*)
    msg.blocking_nodes = all_other_nodes(my_rank); (*@\label{line:other_nodes}@*)
    send(msg);(*@\label{line:send1}@*)
    
    int ret = MPI_BCast(buffer,count,type,root,comm);
    
    msg.state = Running;
    msg.blocking_nodes = 0;
    send(msg);(*@\label{line:send2}@*)
    return ret;
}
\end{lstlisting}
\vspace{-3mm}
\subsubsection{Report Manager}
A report manager is responsible for sending report messages to the power distribution controller. The report manager initially buffers any message until a predefined timeout has passed. Once the timeout expires, the report manager observes the messages queued up in its internal buffer. If a message is followed by another message that cancels it, the report manager skips both messages. For instance, the send call in line~\ref{line:send1} is queued first, and then the send call in line~\ref{line:send2}. If the timeout expires and the socket manager finds both sends in the buffer, it discards both messages. This behavior helps avoid thrashing the CPU frequency and the power distribution controller with frequent and opposing changes. The timeout period is determined using the breakeven solution to the popular ski-rental problem. In this case, the breakeven point is equivalent to the round-trip-time of a report message to be sent to the power distribution controller, and the distribute message to be sent to the affected nodes. If the \code{MPI\_BCast} call ends before the round-trip-time, the report manager avoids thrashing by discarding the message, otherwise it sends the report message. The worst case of the breakeven algorithm is that the \code{MPI\_BCast} call ends immediately at the round-trip-time, in which case sending the report message will not result in any improvement in the performance.
\vspace{-2mm}
\begin{figure}[h]
\centering
\includegraphics[width=0.9\columnwidth]{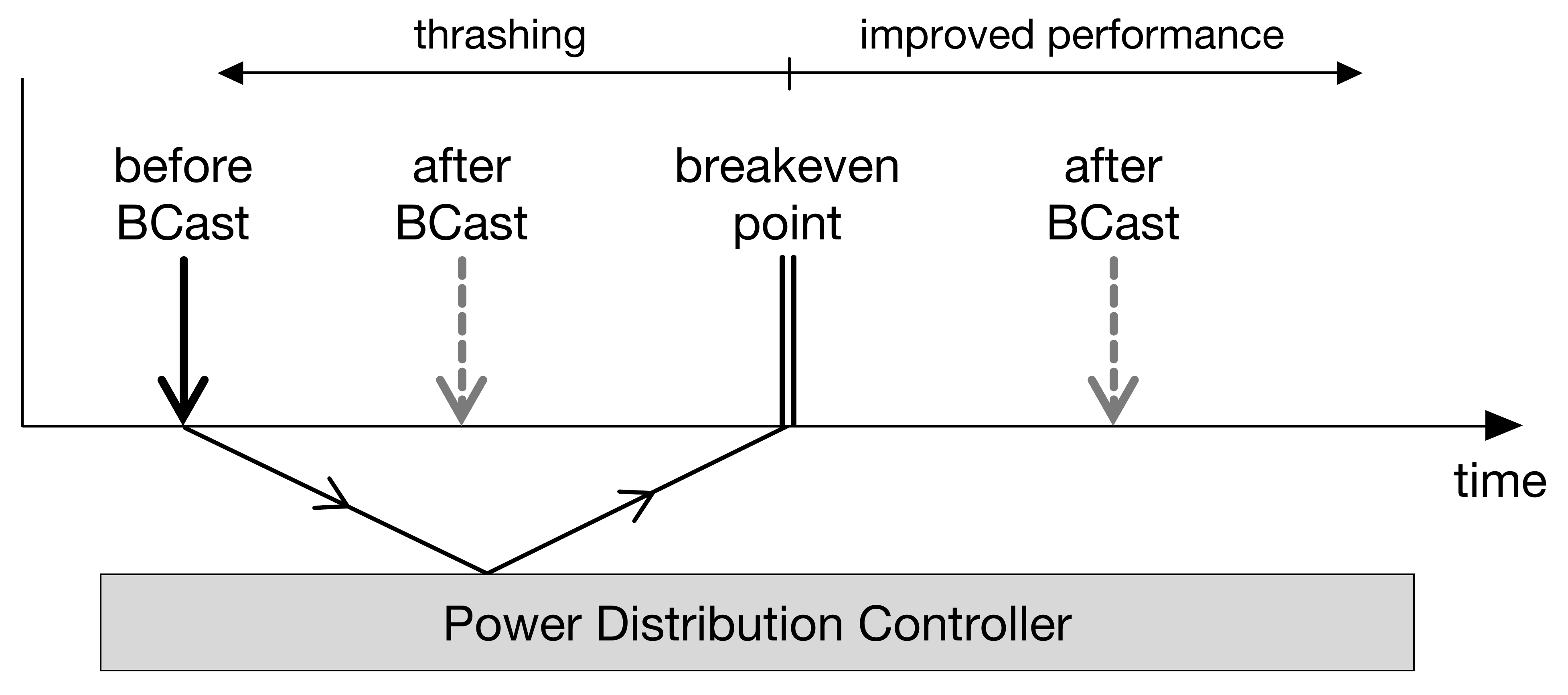}
\caption{The breakeven point at which the report manager checks the buffer.}
\label{fig:breakeven}
\vspace{-3mm}
\end{figure}

The wrapper is missing implementations for \code{MPI\_IRecv} and other asynchronous functions. Also, it lacks support for multiple communicators, which would require a hierarchical approach to power distribution.

\subsection{Experimental setup}
\label{subsec:experiments}

To validate the heuristic in practice, we run $3$ MPI benchmarks on $2$ ARM based boards: (1) Arndale Exynos 5410, hosting a dual-core A15 CPU, and (2) odroid XU-2, hosting a quad-core A15 CPU. ARM has recently gained traction in the HPC domain as a power efficient contender to intel~\cite{rajovic2014tibidabo,armvsintel}. The Arndale board runs linaro ubuntu trusty ($14.04$), while the odroid runs linaro ubuntu raring ($13.04$). This selection of varying manufacturer, CPU capabilities, and OS and kernel versions mimics what would be available at a larger scale in heterogenous clusters. Both boards use OpenMPI 1.8.2, and are connected to an Extech 380803 Power Analyzer that measures their collective power consumption.

%

We run $3$ benchmarks in the NAS Parallel Benchmark suite (NPB). The benchmarks are as follows:
\begin{itemize}
\item \textbf{IS}. An integer sort benchmark that is memory intensive.
\item \textbf{EP}. Embarrassingly parallel benchmark that is CPU intensive.
\item \textbf{CG}. The conjugate gradient benchmark that is communication intensive.
\end{itemize}
For each benchmark, we run three problem size classes: A, B and C. The cluster power bound $\mathbb{P}$ for all experiments is $13$ watts, which is a moderately aggressive power bound given the operating power levels of both boards. We repeat each experiment 3 times to ensure the results are not biased by noise.

\subsection{Experimental Results}

\begin{figure}[h]
\centering
\includegraphics[width=0.8\columnwidth]{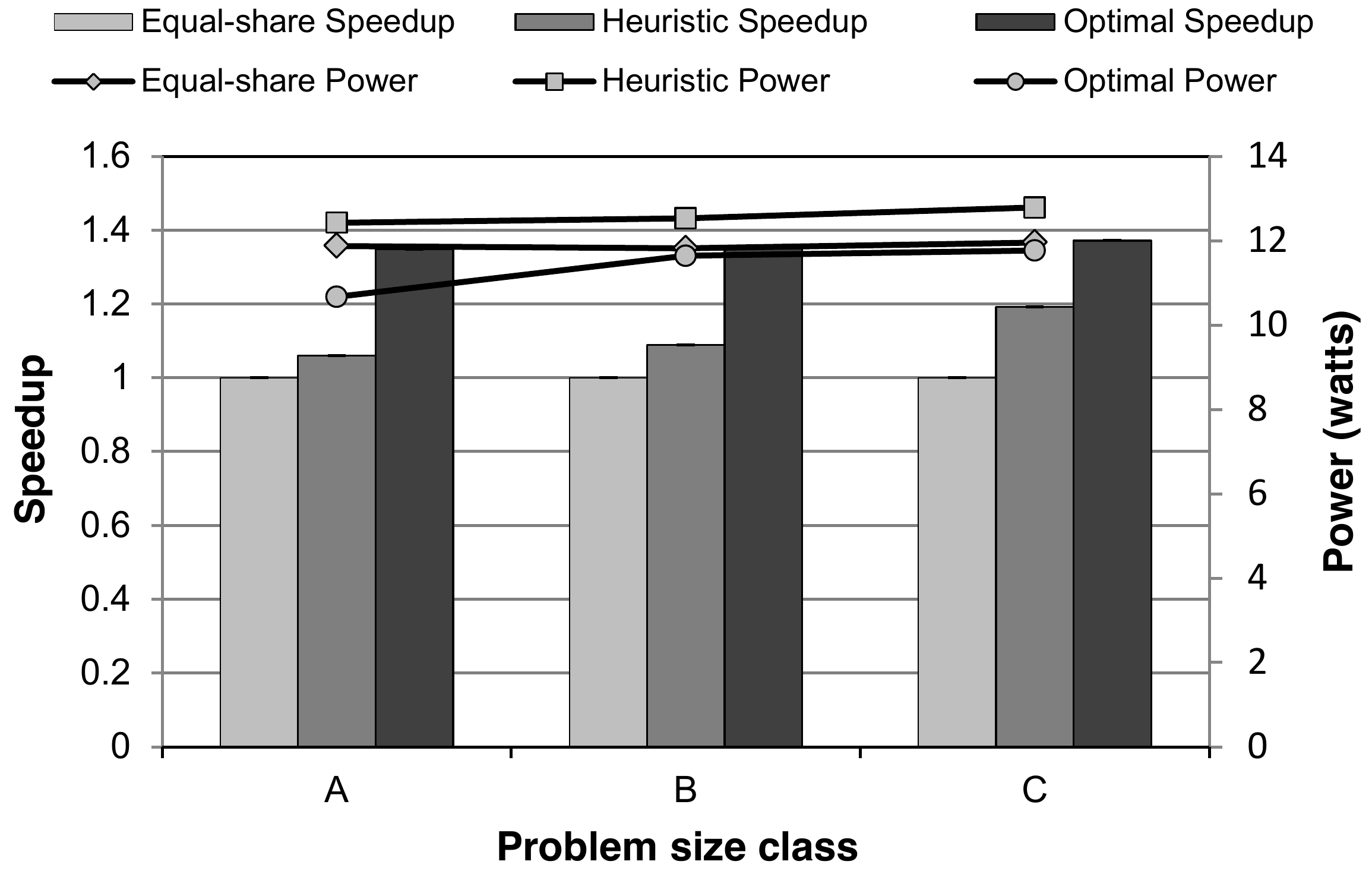}
\caption{Speedup and average power consumption of the IS benchmark.}
\label{fig:is}
\vspace{-5mm}
\end{figure}

\begin{figure}[h]
\centering
\includegraphics[width=0.8\columnwidth]{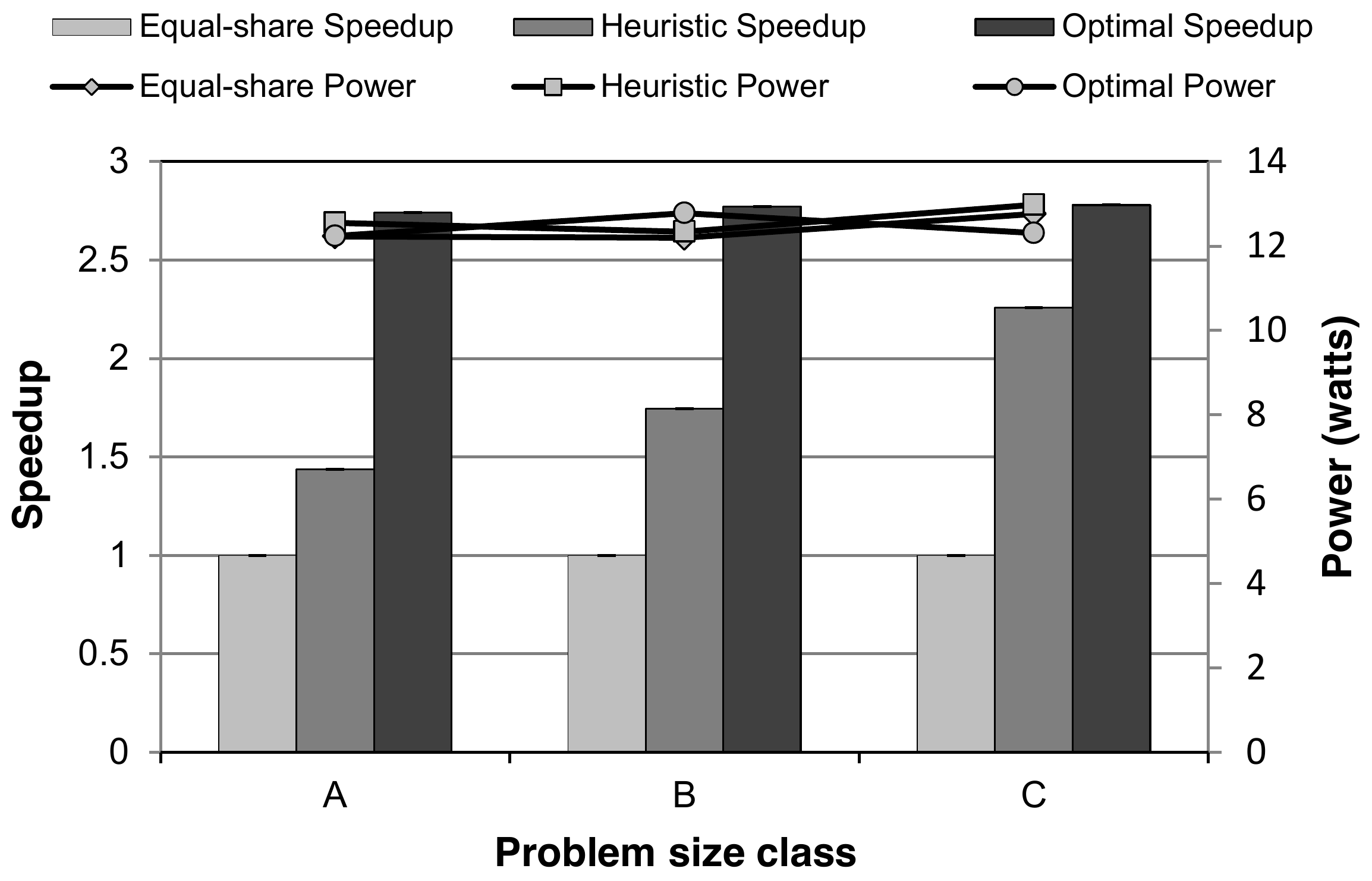}
\caption{Speedup and average power consumption of the EP benchmark.}
\label{fig:ep}
\vspace{-8mm}
\end{figure}

\begin{figure}[h]
\centering
\includegraphics[width=0.8\columnwidth]{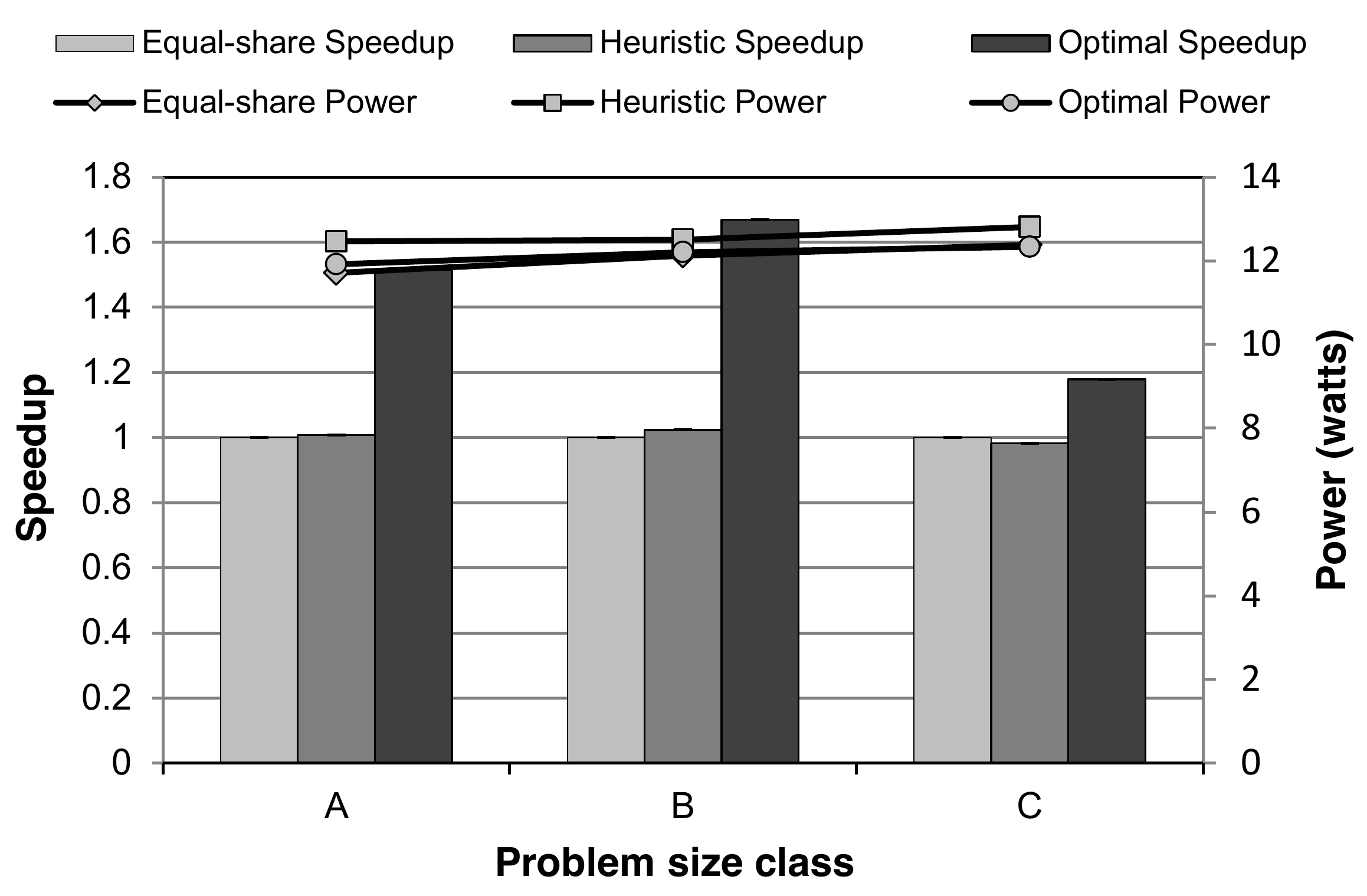}
\caption{Speedup and average power consumption of the CG benchmark.}
\label{fig:cg}
\vspace{-5mm}
\end{figure}

Figures~\ref{fig:is},~\ref{fig:ep}, and~\ref{fig:cg} show the results of executing the IS, EP and CG benchmarks respectively. In the case of IS, the heuristic speed up improves at large problem sizes. This is attributed to the ability of the heuristic to improve performance when the difference in execution time between nodes increases. The power consumption of all three power distribution methods is roughly similar, however, the heuristic power consumption is almost always higher than equal-share or ILP. This observation applies to all three benchmarks, and is attributed to the time discrepancy between a node running after being blocked, and the power distribution controller informing other nodes to lower their power levels to accommodate for the surge that occurs due to the now active node.

The speedup produced by the heuristic is significantly increased in EP, which is expected since the benchmark is heavily CPU bound. At class C, speedup reachers $2.25$, and approaches ILP speedup of $2.78$. This result indicates that the heuristic is better suited for CPU bound MPI programs.

Finally, the heuristic shows inability to improve the CG benchmark. Being communication intensive, the heuristic suffers from two weaknesses: (1) it has very limited time to distribute power more efficiently, rendering it ineffective, and (2) it suffers from some unavoidable thrashing due to the frequency of communication in the program. The interesting point to make here is that the heuristic has minimal {\em negative} effect on performance. Out of the $9$ trials associated with CG, one trial produced a speed-down of $0.98$.  This seemingly scalable stability is attributed to the efficiency of power distribution. Since changing CPU frequency induces minimum overhead versus for instance efficiently distributing workload.

%% file: related.tex
\section{Related Work}
\label{sec:related}

Recent approaches to improve performance of heterogenous clusters rely on applying load-balancing techniques, where the workload is altered to match the computational capabilities of specific nodes~\cite{gandhi2013pikachu}. Our approach does not require redesigning parallel algorithms to support heterogeneous clusters, since the heuristic adapts to variability in execution time whether caused by an unbalanced workload or non-equivalent computational capabilities.

The work in~\cite{meisner2009powernap,gandhi2009optimal} provides a strong foundation on which we base our work. Our approach build on top of this work to introduce a dependency model that can be exploited to improve performance within a power bound.

The work in~\cite{satish2008scheduling} tackles the problem of scheduling dependent jobs on heterogeneous multiprocessors. Our approach attempts to tackle the problem from a power perspective, in the sense that we transfer power dynamically from one node to another upon detecting dependency. This approach helps produce significantly higher speedups than monitoring each node independently. Our implementation of the heuristic requires no knowledge of the deadlines or execution times of workloads, and infers dependency online using parameters of MPI calls.
\vspace{-2mm}

%% file: conclusion.tex
\section{Conclusion}
\label{sec:concl}

In this paper, we tackled the problem of optimizing the performance of MPI 
programs on HPC clusters subject to a power bound. There is little work on 
this problem in the literature, but we argue that given energy constraints of 
HPC clusters and data centers and the increasing demand for computing power, we 
are in pressing need to address the problem. We introduced a formulation of the 
power distribution problem, and presented an ILP-based solution to obtain the 
optimal job-to-power-bound assignment. We then introduced an online heuristic 
that detects when a node is blocked and, subsequently, redistributes its power 
to other nodes based on a ranking algorithm. We validated the approach using 
simulation and actual experiments. Our online heuristic produces a speedup of up 
to a factor of $2.25$, specially in CPU bound programs, while it is ineffective 
in communication intensive applications.

For future work, we planning on running larger experiments on real-sized HPC 
clusters. An interesting research problem is to leverage more information about 
the program using static analysis in an effort to build a smarter heuristic. 
Also, integrating learning mechanisms in the heuristic will allow it to make 
more efficient power distribution decisions.

%% file: main.bbl
\begin{thebibliography}{10}

\bibitem{bailey1992parallel}
David~H Bailey, Leonardo Dagum, Eric Barszcz, and Horst~D Simon.
\newblock Nas parallel benchmark results.
\newblock In {\em Proceedings of the 1992 ACM/IEEE conference on
  Supercomputing}, pages 386--393. IEEE Computer Society Press, 1992.

\bibitem{barroso2005price}
Luiz~Andr{\'e} Barroso.
\newblock The price of performance.
\newblock {\em Queue}, 3(7):48--53, 2005.

\bibitem{armvsintel}
Jeffrey Burt.
\newblock {Intel, ARM take competition into HPC arena}.
\newblock
  \url{http://www.eweek.com/servers/intel-arm-take-competition-into-hpc-arena.html},
  2014.
\newblock [Online; accessed 18-October-2014].

\bibitem{gandhi2009optimal}
Anshul Gandhi, Mor Harchol-Balter, Rajarshi Das, and Charles Lefurgy.
\newblock Optimal power allocation in server farms.
\newblock In {\em ACM SIGMETRICS Performance Evaluation Review}, volume~37,
  pages 157--168. ACM, 2009.

\bibitem{gandhi2013pikachu}
Rohan Gandhi, Di~Xie, and Y~Charlie Hu.
\newblock Pikachu: How to rebalance load in optimizing mapreduce on
  heterogeneous clusters.
\newblock In {\em USENIX Annual Technical Conference}, pages 61--66, 2013.

\bibitem{heath2005energy}
Taliver Heath, Bruno Diniz, Enrique~V Carrera, Wagner Meira~Jr, and Ricardo
  Bianchini.
\newblock Energy conservation in heterogeneous server clusters.
\newblock In {\em Proceedings of the tenth ACM SIGPLAN symposium on Principles
  and practice of parallel programming}, pages 186--195. ACM, 2005.

\bibitem{koomey2011growth}
Jonathan Koomey.
\newblock Growth in data center electricity use 2005 to 2010.
\newblock {\em A report by Analytical Press, completed at the request of The
  New York Times}, 2011.

\bibitem{meisner2009powernap}
David Meisner, Brian~T Gold, and Thomas~F Wenisch.
\newblock Powernap: eliminating server idle power.
\newblock In {\em ACM Sigplan Notices}, volume~44, pages 205--216. ACM, 2009.

\bibitem{rajovic2014tibidabo}
Nikola Rajovic, Alejandro Rico, Nikola Puzovic, Chris Adeniyi-Jones, and Alex
  Ramirez.
\newblock Tibidabo: Making the case for an arm-based hpc system.
\newblock {\em Future Generation Computer Systems}, 36:322--334, 2014.

\bibitem{ramapantulu2014modeling}
Lavanya Ramapantulu, Bogdan~Marius Tudor, Dumitrel Loghin, Trang Vu, and
  Yong~Meng Teo.
\newblock Modeling the energy efficiency of heterogeneous clusters.
\newblock ICPP, 2014.

\bibitem{satish2008scheduling}
Nadathur~R Satish, Kaushik Ravindran, and Kurt Keutzer.
\newblock Scheduling task dependence graphs with variable task execution times
  onto heterogeneous multiprocessors.
\newblock In {\em Proceedings of the 8th ACM international conference on
  Embedded software}, pages 149--158. ACM, 2008.

\end{thebibliography}
